\documentclass[10pt,reqno]{amsart}
\usepackage[T1]{fontenc}
\usepackage[latin9]{inputenc}
\usepackage{amssymb}
\usepackage{esint}
\usepackage{fullpage}

 \usepackage[foot]{amsaddr}



\title{Revised identification of strain gradient elastic parameters}

\author[1]{Luca Placidi$^1$}
\author[2]{Anil Misra$^2$}
\author[3]{Gabriele La Valle$^3$}
\author[4]{Casey Rodriguez$^4$}
\address{$^1$Department of Engineering, Telematic University Pegaso Centro
	Direzionale ISOLA F2, Napoli, Italy}
\address{$^2$Florida International
	University, Miami, Florida, USA}
\address{$^3$Laboratory MSME, UMR 8208 Université Gustave Eiffel, France}
\address{$^4$The University of
	North Carolina at Chapel Hill, Chapel Hill, NC, USA}

\begin{document}
\begin{abstract}
The work reported in ``Granular micromechanics-based identification of isotropic strain gradient parameters for elastic geometrically nonlinear deformations" misidentified key terms in the grain-pair objective relative displacement when accounting for the second gradient of placement. In this paper, we correct that oversight by deriving a revised expression for the grain-pair objective relative displacement within the granular micromechanics framework. The amended terms, which resemble Christoffel symbols expressed in terms of strain gradients, modify the contributions of both the normal and tangential components to the strain energy and, consequently, alter the identified strain gradient elastic parameters. Importantly, the identification of the standard (first gradient) elastic tensor remains unchanged. This brief paper presents the corrected derivation, the resulting stiffness tensors for anisotropic strain gradient elasticity, and updated analytical expressions for the material parameters in both 2D and 3D isotropic settings. 
\end{abstract}

\maketitle

\section{Introduction}

In recent years, strain gradient elasticity theories have garnered
significant attention for their ability to capture size-dependent
mechanical behavior in nano-structured materials and metamaterials,
where classical continuum models fail to provide accurate predictions
\cite{Eugster_et_al_2022,Eremeyev_et_al_2020,Barchiesi_et_Al_2024,giorgio_Et_al_2024,Turco_et_al_2022}.
These advanced models, particularly those incorporating higher-order
or second strain gradient effects, have been successfully applied
to investigate a wide range of phenomena, such as wave propagation
in periodic nanostructures and graphene sheets \cite{yang_et_al_2024_cs,Yang_et_al_2024_tws,Yang_et_al_2023},
static and dynamic analyses of nanoplates and nanobeams \cite{Saitta_et_al_2022_ijcm,Saitta_et_al_2022_eabe,Fabbrocino_et_al_2022,Cornacchia_et_al_2021,Tocci_Monaco_et_al_2021,Tocci_Monaco_et_al_2021_cs,Penna_Et_al_2021,Tocci_Monaco_et_al_2021_m,Tocci_Monaco_et_al_2021_n},
and hygro-thermal responses of functionally graded and magneto-electro-elastic
nano-components \cite{Tocci_Monaco_et_al_2021_cs,Penna_Et_al_2021,Tocci_Monaco_et_al_2021_m,Tocci_Monaco_et_al_2021_n}.
Among these, strain gradient elasticity---particularly its higher-order
and second-gradient formulations---has proven effective in describing
phenomena such as wave dispersion, localization, and boundary layer
effects in nano-beams, plates, and metamaterials \cite{Ussorio_et_al_2025,Barretta_etA_l_2023,Darban_et_al_2022,Caporale_Et_al_2022,Bacciocchi_et_al_2021,Apuzzo_et_al_2019,Apuzzo_et_al_2018,Barretta_et_al_2018}.
Computational approaches leveraging meshless methods \cite{Saitta_et_al_2022_ijcm,Saitta_et_al_2022_eabe,Fabbrocino_et_al_2022},
finite element techniques \cite{Bacciocchi_et_al_2023,Bacciocchi_et_al_2021},
and semi-analytical formulations \cite{Tocci_Monaco_et_al_2021_m}
have further enhanced the applicability of strain gradient models
to complex geometries and loading conditions. Moreover, recent developments
in homogenization techniques and metamaterial modeling have emphasized
the role of second gradient continua in capturing microscale interactions
and deformation patterns in engineered structures \cite{Ciallella_et_al_2023,Yang_H_et_al_2023,Stilz_et_al_2022,McAvoy_barchiesi_2025,Barchiesi_2024,Turco_Et_al_2024,Ciallella_Et_al_2024}.
Despite this growing body of work, the accurate identification of
strain gradient elastic coefficients---crucial for both predictive
modeling and material design---remains an open challenge. In this
context, the present study aims to contribute to this field by proposing
a robust framework for the identification of such coefficients, thereby
advancing the reliability and effectiveness of strain gradient theories
in engineering applications. In other words, a central challenge in
deploying these models lies in the accurate identification of the
additional constitutive parameters introduced by gradient elasticity.
While most existing works focus on the implementation of gradient
theories in numerical and analytical frameworks, including finite
element and meshless methods \cite{Darban_et_al_2022,Caporale_Et_al_2022,Reiher_Et_al_2017,Giorgio_2016},
only a limited number of studies address the direct physical interpretation
and quantitative determination of the gradient moduli. Recent advances,
particularly those based on homogenization \cite{Rahali_et_al_2015,Rosi_et_al_2013,Fallahnejad_et_al_2023,Barchiesi_Et_al_2023_ejms}
and microstructure-informed modeling \cite{Danesh_et_al_2023,Barchiesi_et_al_2023,dell_Isola_et_al_2024,dell_Isola_et_al_2023,dell_Isola_et_al_2022},
have paved the way toward more physically grounded identification
techniques. Further developments in this direction include the use
of multiscale and stress-driven models \cite{Shekarchizadeh_Et_al_2021,Tran_barchiesi_2023,Solyaev_et_al_2022,Eugster_et_al_2022,Eremeyev_et_al_2020},
as well as numerical studies of lattice and pantographic structures
\cite{McAvoy_barchiesi_2025,Barchiesi_2024,Turco_Et_al_2024,Ciallella_Et_al_2024}.
Building on these foundations, the present study introduces a granular
micromechanical approach for the identification of strain gradient
elastic moduli, grounded in the physical interpretation of microscale
interactions and compatible with continuum models of second-gradient
type. This contribution seeks to bridge the gap between micromechanical
modeling and constitutive parameter identification, enabling more
predictive and physically consistent applications of strain gradient
elasticity.

\section{A recap before the correction}

In granular micromechanics, the
density of the (internal) strain energy is, according to (15) of \cite{barchiesietlZAMM2021},
the integral of the energy of each orientation $\hat{c}$ over the
unit circle $S^{\mathrm{1}}$ in the 2D case or over the unit sphere
$S^{2}$ in the 3D case,
\begin{equation}
U=\int_{\mathcal{S^{\mathrm{1,2}}}}\left(\frac{1}{2}k_{\eta}u_{\eta}^{2}+\frac{1}{2}k_{\tau}u_{\tau}^{2}\right)dS,\label{eq:density elastic energy-1}
\end{equation}
where $k_{\eta}$ is the normal stiffness relative to the orientation
$\hat{c}$, $k_{\tau}$ is the tangent stiffness relative to the orientation
$\hat{c}$, and $dS$ is the angular element over $S^{\mathrm{1}}$
or $S^{2}$. In addition, $u_{\eta}$ is the normal displacement, which
is a scalar, and $u_{\tau}$ is the tangent displacement, which is
a vector. The normal displacement is defined by equation (11) of \cite{barchiesietlZAMM2021},
\begin{equation}
u_{\eta}=\frac{1}{2}u^{np}\cdot\hat{c}=\frac{1}{2}u_{i}^{np}\hat{c}_{i},\label{eq:ueta1}
\end{equation}
and the tangential displacement by the equation (12) of \cite{barchiesietlZAMM2021},
\begin{equation}
u_{\tau}=u^{np}-\left(u^{np}\cdot\hat{c}\right)\hat{c}=u^{np}-2u_{\eta}\hat{c}.\label{eq:utau1}
\end{equation}
In (\ref{eq:ueta1}-\ref{eq:utau1}) we use the definition of the
objective relative displacement, $u^{np}$, which is defined below (see \eqref{eq:objreldispl}). 

\section{The objective relative displacement and its projections}

The objective relative displacement is, from the first equivalence
of the equation (9) of \cite{barchiesietlZAMM2021},
\begin{equation}
u^{np}=2G\hat{c}L+\frac{L^{2}}{2}\left(F^{T}\nabla F\right)\hat{c}\hat{c},\label{eq:objreldispl}
\end{equation}
where $G$ is the Green-Saint-Venant strain tensor,
\begin{equation}
G=\frac{1}{2}\left(F^{T}F-I\right),\label{eq:green saint venant}
\end{equation}
$F$ is the deformation gradient,
\begin{equation}
F=\nabla\chi,\label{eq:deformation gradiet}
\end{equation}
$\chi$ is the placement function, $\hat{c}$ is a unit vector, $L$
is average distance between the neighboring grain-pairs and $\nabla$
is the gradient operator. In index notation (\ref{eq:objreldispl})
yields

\begin{equation}
u_{i}^{np}=2G_{ij}\hat{c}_{j}L+\frac{L^{2}}{2}F_{ia}^{T}\left(\nabla F\right)_{abc}\hat{c}_{c}\hat{c}_{b}=2G_{ij}\hat{c}_{j}L+\frac{L^{2}}{2}H_{ibc}\hat{c}_{c}\hat{c}_{b},\label{eq:objecti relative displ index}
\end{equation}
where the definition (\ref{eq:green saint venant}) has been taken
into account.
\[
H_{ibc}=F_{ai}F_{ab,c}=\chi_{a,i}\chi_{a,bc}=H_{icb}
\]
is a third order tensor, symmetric with respect the last two indices.
The error of equation (9) in \cite{barchiesietlZAMM2021} is due to
a wrong identification of the last term of (\ref{eq:objecti relative displ index}),
\[
\frac{L^{2}}{2}H_{ibc}\hat{c}_{c}\hat{c}_{b}\rightarrow\frac{L^{2}}{2}G_{ib,c}\hat{c}_{c}\hat{c}_{b},
\]
which is simply an oversight. Indeed, it is possible to
prove the following identity for the components of the tensor $H$, similar in form to 
Christoffel symbols:
\begin{equation}
H_{ibc}=G_{ib,c}+G_{ic,b}-G_{bc,i}.\label{eq:H  definition}
\end{equation}
Let us prove (\ref{eq:H  definition}). From (\ref{eq:green saint venant})
we have
\[
G_{ij}=\frac{1}{2}\left(F_{ki}F_{kj}-\delta_{ij}\right).
\]
Its gradient is
\begin{equation}
G_{ij,h}=\frac{1}{2}\left(F_{ki,h}F_{kj}+F_{ki}F_{kj,h}\right).\label{eq:gradient G indece}
\end{equation}
Thus, by changing the names of the indices of (\ref{eq:gradient G indece})
we have the following three expressions,
\begin{align*}
 G_{ib,c}&=\frac{1}{2}\left(F_{ki,c}F_{kb}+F_{ki}F_{kb,c}\right),\\
 G_{ic,b}&=\frac{1}{2}\left(F_{ki,b}F_{kc}+F_{ki}F_{kc,b}\right),\\
 G_{bc,i}&=\frac{1}{2}\left(F_{kb,i}F_{kc}+F_{kb}F_{kc,i}\right).
\end{align*}
From the previous three equations, the combination expressed on th
right-hand side of (\ref{eq:H  definition}) yields
\begin{equation}
G_{ib,c}+G_{ic,b}-G_{bc,i}=\frac{1}{2}\left(F_{ki,c}F_{kb}+F_{ki}F_{kb,c}+F_{ki,b}F_{kc}+F_{ki}F_{kc,b}-F_{kb,i}F_{kc}-F_{kb}F_{kc,i}\right).\label{eq:onter}
\end{equation}
Because of the symmetry on the last two indices of the gradient of
the deformation gradient in (\ref{eq:deformation gradiet}), 
\[
\left(\nabla F\right)_{abc}=F_{ab,c}=\chi_{a,bc}=\chi_{a,cb}=F_{ac,b}=\left(\nabla F\right)_{acb},
\]
the last two terms of (\ref{eq:onter}) cancel the first, and the
third terms and the second term equals the fourth one. As a consequence,
we have,
\[
G_{ib,c}+G_{ic,b}-G_{bc,i}=F_{ki}F_{kb,c}=H_{ibc}=H_{icb},
\]
proving (\ref{eq:H  definition}).

Thus, the objective relative displacement is given by the insertion
of (\ref{eq:H  definition}) into (\ref{eq:objecti relative displ index})
\begin{equation}
u_{i}^{np}=2G_{ij}\hat{c}_{j}L+\frac{L^{2}}{2}\left(G_{ib,c}+G_{ic,b}-G_{bc,i}\right)\hat{c}_{c}\hat{c}_{b}.\label{eq:objective relative displacement index}
\end{equation}
The normal displacement $u_{\eta}$ is defined by equation (11) of
\cite{barchiesietlZAMM2021}. The insertion of (\ref{eq:objective relative displacement index})
into the definition yields,
\begin{equation}
u_{\eta}=\frac{1}{2}u_{i}^{np}\hat{c}_{i}=G_{ij}\hat{c}_{i}\hat{c}_{j}L+\frac{L^{2}}{4}\left(G_{ib,c}\hat{c}_{i}\hat{c}_{c}\hat{c}_{b}+G_{ic,b}\hat{c}_{i}\hat{c}_{c}\hat{c}_{b}-G_{bc,i}\hat{c}_{i}\hat{c}_{c}\hat{c}_{b}\right).\label{eq:normal displacement}
\end{equation}
The first and the third terms in parenthesis are the same because
of symmetry (exchanging the names of the indices $i$ and $b$), so
that the normal displacement (\ref{eq:normal displacement}) is
\[
u_{\eta}=G_{ij}\hat{c}_{i}\hat{c}_{j}L+\frac{L^{2}}{4}G_{ij,h}\hat{c}_{i}\hat{c}_{j}\hat{c}_{h},
\]
the same expression as (A.2) of the \cite{barchiesietlZAMM2021}.
Its square is
\[
u_{\eta}^{2}=\left(G_{ij}\hat{c}_{i}\hat{c}_{j}L+\frac{L^{2}}{4}G_{ij,h}\hat{c}_{i}\hat{c}_{j}\hat{c}_{h}\right)\left(G_{ab}\hat{c}_{a}\hat{c}_{b}L+\frac{L^{2}}{4}G_{ab,c}\hat{c}_{a}\hat{c}_{b}\hat{c}_{c}\right).
\]
Expanding the square of the previous binomial
\begin{align*}
u_{\eta}^{2}= &  \left(G_{ij}\hat{c}_{i}\hat{c}_{j}L\right)\left(G_{ab}\hat{c}_{a}\hat{c}_{b}L\right)\\
 &   +\left(G_{ij}\hat{c}_{i}\hat{c}_{j}L\right)\left(\frac{L^{2}}{4}G_{ab,c}\hat{c}_{a}\hat{c}_{b}\hat{c}_{c}\right)\\
 &   +\left(\frac{L^{2}}{4}G_{ij,h}\hat{c}_{i}\hat{c}_{j}\hat{c}_{h}\right)\left(G_{ab}\hat{c}_{a}\hat{c}_{b}L\right)\\
 &   +\left(\frac{L^{2}}{4}G_{ij,h}\hat{c}_{i}\hat{c}_{j}\hat{c}_{h}\right)\left(\frac{L^{2}}{4}G_{ab,c}\hat{c}_{a}\hat{c}_{b}\hat{c}_{c}\right),
\end{align*}
we have the equivalence between the second and the third lines of
the previous equation. Thus,
\begin{equation}
u_{\eta}^{2}=L^{2}G_{ij}G_{ab}\hat{c}_{i}\hat{c}_{j}\hat{c}_{a}\hat{c}_{b}+\frac{L^{3}}{2}G_{ij}G_{ab,c}\hat{c}_{i}\hat{c}_{j}\hat{c}_{a}\hat{c}_{b}\hat{c}_{c}+\frac{L^{4}}{16}G_{ij,h}G_{ab,c}\hat{c}_{i}\hat{c}_{j}\hat{c}_{h}\hat{c}_{a}\hat{c}_{b}\hat{c}_{c}.\label{eq:squared noormal displacement}
\end{equation}
The tangent displacement is a vector, that is defined in equation
(12) of \cite{barchiesietlZAMM2021}, 
\[
u_{\tau}=u^{np}-\left(u^{np}\cdot\hat{c}\right)\hat{c}=u^{np}-2u_{\eta}\hat{c},
\]
where we have used the definition (\ref{eq:normal displacement}).
In index notation the previous expression is 
$
u_{\tau i}=u_{i}^{np}-2u_{\eta}\hat{c}_{i}.
$
The squared tangent displacement is the scalar
\begin{equation}
u_{\tau}^{2}=\left(u_{i}^{np}-2u_{\eta}\hat{c}_{i}\right)\left(u_{i}^{np}-2u_{\eta}\hat{c}_{i}\right)=u_{i}^{np}u_{i}^{np}-4u_{\eta}u_{i}^{np}\hat{c}_{i}+4u_{\eta}^{2}=u_{i}^{np}u_{i}^{np}-4u_{\eta}^{2}.\label{eq:squared tangent displacement}
\end{equation}
Let us analyze the first term of (\ref{eq:squared tangent displacement}),
which is the norm of the objective relative displacement,
\begin{align}
 u_{i}^{np}u_{i}^{np}=&\left[2G_{ib}\hat{c}_{b}L+\frac{L^{2}}{2}\left(G_{ib,c}+G_{ic,b}-G_{bc,i}\right)\hat{c}_{c}\hat{c}_{b}\right]\label{eq:squared objective relative displacement}\\
 & \times\left[2G_{ij}\hat{c}_{j}L+\frac{L^{2}}{2}\left(G_{ij,h}+G_{ih,j}-G_{jh,i}\right)\hat{c}_{h}\hat{c}_{j}\right].\nonumber 
\end{align}
Expanding the square of the previous binomial we have
\begin{align}
u_{i}^{np}u_{i}^{np}= &   \left[2G_{ib}\hat{c}_{b}L\right]\left[2G_{ij}\hat{c}_{j}L\right]\\
 &   +\left[2G_{ib}\hat{c}_{b}L\right]\left[\frac{L^{2}}{2}\left(G_{ij,h}+G_{ih,j}-G_{jh,i}\right)\hat{c}_{h}\hat{c}_{j}\right]\\
 &   +\left[\frac{L^{2}}{2}\left(G_{ib,c}+G_{ic,b}-G_{bc,i}\right)\hat{c}_{c}\hat{c}_{b}\right]\left[2G_{ij}\hat{c}_{j}L\right]\\
 &   +\left[\frac{L^{2}}{2}\left(G_{ib,c}+G_{ic,b}-G_{bc,i}\right)\hat{c}_{c}\hat{c}_{b}\right]\left[\frac{L^{2}}{2}\left(G_{ij,h}+G_{ih,j}-G_{jh,i}\right)\hat{c}_{h}\hat{c}_{j}\right].
\end{align}
The second and the third lines of the previous expression are the
same. Thus the norm of the objective relative displacement (\ref{eq:squared objective relative displacement})
is,
\begin{align*}
u_{i}^{np}u_{i}^{np} &=  4L^{2}G_{ib}G_{ij}\hat{c}_{b}\hat{c}_{j}\\
 &   +2L^{3}G_{ib}\left(G_{ij,h}+G_{ih,j}-G_{jh,i}\right)\hat{c}_{b}\hat{c}_{h}\hat{c}_{j}\\
 &   +\frac{L^{4}}{4}\left(G_{ib,c}+G_{ic,b}-G_{bc,i}\right)\left(G_{ij,h}+G_{ih,j}-G_{jh,i}\right)\hat{c}_{c}\hat{c}_{b}\hat{c}_{h}\hat{c}_{j}.
\end{align*}
Expanding each term of the previous expressions we have
\begin{align*}
u_{i}^{np}u_{i}^{np} &= 4L^{2}G_{ib}G_{ij}\hat{c}_{b}\hat{c}_{j}\\
 &   +2L^{3}G_{ib}G_{ij,h}\hat{c}_{b}\hat{c}_{h}\hat{c}_{j}\\
 &   +2L^{3}G_{ib}G_{ih,j}\hat{c}_{b}\hat{c}_{h}\hat{c}_{j}\\
 &   -2L^{3}G_{ib}G_{jh,i}\hat{c}_{b}\hat{c}_{h}\hat{c}_{j}\\
 &   +\frac{L^{4}}{4}G_{ib,c}G_{ij,h}\hat{c}_{c}\hat{c}_{b}\hat{c}_{h}\hat{c}_{j}\\
 &   +\frac{L^{4}}{4}G_{ib,c}G_{ih,j}\hat{c}_{c}\hat{c}_{b}\hat{c}_{h}\hat{c}_{j}\\
 &   -\frac{L^{4}}{4}G_{ib,c}G_{jh,i}\hat{c}_{c}\hat{c}_{b}\hat{c}_{h}\hat{c}_{j}\\
 &  +\frac{L^{4}}{4}G_{ic,b}G_{ij,h}\hat{c}_{c}\hat{c}_{b}\hat{c}_{h}\hat{c}_{j}\\
 &   +\frac{L^{4}}{4}G_{ic,b}G_{ih,j}\hat{c}_{c}\hat{c}_{b}\hat{c}_{h}\hat{c}_{j}\\
 &   -\frac{L^{4}}{4}G_{ic,b}G_{jh,i}\hat{c}_{c}\hat{c}_{b}\hat{c}_{h}\hat{c}_{j}\\
 &   -\frac{L^{4}}{4}G_{bc,i}G_{ij,h}\hat{c}_{c}\hat{c}_{b}\hat{c}_{h}\hat{c}_{j}\\
 &   -\frac{L^{4}}{4}G_{bc,i}G_{ih,j}\hat{c}_{c}\hat{c}_{b}\hat{c}_{h}\hat{c}_{j}\\
 &   +\frac{L^{4}}{4}G_{bc,i}G_{jh,i}\hat{c}_{c}\hat{c}_{b}\hat{c}_{h}\hat{c}_{j}.
\end{align*}
Changing the names of the indices over which we have an implicit sum
in order to collect the same strain and strain gradient components,
the previous expression is
\begin{align*}
u_{i}^{np}u_{i}^{np} &= 4L^{2}G_{ab}G_{ij}\delta_{ia}\hat{c}_{b}\hat{c}_{j}\\
 &   +2L^{3}G_{ab}G_{ij,h}\delta_{ia}\hat{c}_{b}\hat{c}_{h}\hat{c}_{j}\\
 &   +2L^{3}G_{ab}G_{ij,h}\delta_{ia}\hat{c}_{b}\hat{c}_{h}\hat{c}_{j}\\
 &   -2L^{3}G_{ab}G_{ij,h}\delta_{ha}\hat{c}_{b}\hat{c}_{j}\hat{c}_{i}\\
 &   +\frac{L^{4}}{4}G_{ab,c}G_{ij,h}\delta_{ia}\hat{c}_{c}\hat{c}_{b}\hat{c}_{h}\hat{c}_{j}\\
 &   +\frac{L^{4}}{4}G_{ab,c}G_{ij,h}\delta_{ia}\hat{c}_{c}\hat{c}_{b}\hat{c}_{h}\hat{c}_{j}\\
 &   -\frac{L^{4}}{4}G_{ab,c}G_{ij,h}\delta_{ha}\hat{c}_{c}\hat{c}_{b}\hat{c}_{j}\hat{c}_{i}\\
 &   +\frac{L^{4}}{4}G_{ab,c}G_{ij,h}\delta_{ia}\hat{c}_{c}\hat{c}_{b}\hat{c}_{h}\hat{c}_{j}\\
 &   +\frac{L^{4}}{4}G_{ab,c}G_{ij,h}\delta_{ia}\hat{c}_{c}\hat{c}_{b}\hat{c}_{h}\hat{c}_{j}\\
 &   -\frac{L^{4}}{4}G_{ab,c}G_{ij,h}\delta_{ha}\hat{c}_{c}\hat{c}_{b}\hat{c}_{j}\hat{c}_{i}\\
 &   -\frac{L^{4}}{4}G_{ab,c}G_{ij,h}\delta_{ic}\hat{c}_{b}\hat{c}_{a}\hat{c}_{h}\hat{c}_{j}\\
 &   -\frac{L^{4}}{4}G_{ab,c}G_{ij,h}\delta_{ic}\hat{c}_{b}\hat{c}_{a}\hat{c}_{h}\hat{c}_{j}\\
 &   +\frac{L^{4}}{4}G_{ab,c}G_{ij,h}\delta_{hc}\hat{c}_{b}\hat{c}_{a}\hat{c}_{j}\hat{c}_{i}.
\end{align*}
Let us now collect the terms of the previous expression into three
parts. The first with the squared strain components, the third with
the square of the strain gradient components and the second with the
mixed strain-strain gradient components,
\begin{align}
u_{i}^{np}u_{i}^{np}  &= 4L^{2}G_{ab}G_{ij}\delta_{ia}\hat{c}_{b}\hat{c}_{j}\label{eq:squared objetive relative displacement}\\
 &   +2L^{3}G_{ab}G_{ij,h}\left(2\delta_{ia}\hat{c}_{b}\hat{c}_{h}\hat{c}_{j}-\delta_{ha}\hat{c}_{b}\hat{c}_{j}\hat{c}_{i}\right)\nonumber \\
 &   +\frac{L^{4}}{4}G_{ab,c}G_{ij,h}\left(4\delta_{ia}\hat{c}_{c}\hat{c}_{b}\hat{c}_{h}\hat{c}_{j}-2\delta_{ha}\hat{c}_{c}\hat{c}_{b}\hat{c}_{j}\hat{c}_{i}-2\delta_{ic}\hat{c}_{b}\hat{c}_{a}\hat{c}_{h}\hat{c}_{j}+\delta_{hc}\hat{c}_{b}\hat{c}_{a}\hat{c}_{j}\hat{c}_{i}\right).\nonumber 
\end{align}
Thus, the squared tangent displacement is from the insertion of (\ref{eq:squared noormal displacement})
and (\ref{eq:squared objetive relative displacement}) into (\ref{eq:squared tangent displacement}),
\begin{align*}
u_{\tau}^{2}=u_{i}^{np}u_{i}^{np}-4u_{\eta}^{2}  &  =4L^{2}G_{ab}G_{ij}\left(\delta_{ia}\hat{c}_{b}\hat{c}_{j}-\hat{c}_{i}\hat{c}_{j}\hat{c}_{a}\hat{c}_{b}\right)\\
 &   +2L^{3}G_{ab}G_{ij,h}\left(2\delta_{ia}\hat{c}_{b}\hat{c}_{h}\hat{c}_{j}-\delta_{ha}\hat{c}_{b}\hat{c}_{j}\hat{c}_{i}-\hat{c}_{i}\hat{c}_{j}\hat{c}_{a}\hat{c}_{b}\hat{c}_{c}\right)\\
 &   +\frac{L^{4}}{4}G_{ab,c}G_{ij,h}\left(4\delta_{ia}\hat{c}_{c}\hat{c}_{b}\hat{c}_{h}\hat{c}_{j}-2\delta_{ha}\hat{c}_{c}\hat{c}_{b}\hat{c}_{j}\hat{c}_{i}-2\delta_{ic}\hat{c}_{b}\hat{c}_{a}\hat{c}_{h}\hat{c}_{j}\right)\\
 &   +\frac{L^{4}}{4}G_{ab,c}G_{ij,h}\left(\delta_{hc}\hat{c}_{b}\hat{c}_{a}\hat{c}_{j}\hat{c}_{i}-\hat{c}_{i}\hat{c}_{j}\hat{c}_{h}\hat{c}_{a}\hat{c}_{b}\hat{c}_{c}\right).
\end{align*}
Because of the symmetry of both the strain tensor $G$ and its gradient
$\nabla G$ with respect to the first two indices, we symmetrize the
coefficients of the strain and strain gradient components as follows,
\begin{align*}
u_{\tau}^{2}=u_{i}^{np}u_{i}^{np}-4u_{\eta}^{2}  & =L^{2}G_{ab}G_{ij}\left(\delta_{ia}\hat{c}_{b}\hat{c}_{j}+\delta_{ja}\hat{c}_{b}\hat{c}_{i}+\delta_{ib}\hat{c}_{a}\hat{c}_{j}+\delta_{jb}\hat{c}_{a}\hat{c}_{i}-4\hat{c}_{i}\hat{c}_{j}\hat{c}_{a}\hat{c}_{b}\right)\\
 &   +\frac{L^{3}}{2}G_{ab}G_{ij,h}\left(2\delta_{ia}\hat{c}_{b}\hat{c}_{h}\hat{c}_{j}+2\delta_{ja}\hat{c}_{b}\hat{c}_{h}\hat{c}_{i}+2\delta_{ib}\hat{c}_{a}\hat{c}_{h}\hat{c}_{j}+2\delta_{jb}\hat{c}_{a}\hat{c}_{h}\hat{c}_{i}\right)\\
 &   -\frac{L^{3}}{2}G_{ab}G_{ij,h}\left(2\delta_{ha}\hat{c}_{b}\hat{c}_{j}\hat{c}_{i}+2\delta_{hb}\hat{c}_{a}\hat{c}_{j}\hat{c}_{i}+4\hat{c}_{i}\hat{c}_{j}\hat{c}_{a}\hat{c}_{b}\hat{c}_{c}\right)\\
 &   +\frac{L^{4}}{4}G_{ab,c}G_{ij,h}\left(\delta_{ia}\hat{c}_{c}\hat{c}_{b}\hat{c}_{h}\hat{c}_{j}+\delta_{ib}\hat{c}_{c}\hat{c}_{a}\hat{c}_{h}\hat{c}_{j}+\delta_{ja}\hat{c}_{c}\hat{c}_{b}\hat{c}_{h}\hat{c}_{i}+\delta_{jb}\hat{c}_{c}\hat{c}_{a}\hat{c}_{h}\hat{c}_{i}\right)\\
 &   -\frac{L^{4}}{4}G_{ab,c}G_{ij,h}\left(\delta_{ha}\hat{c}_{c}\hat{c}_{b}\hat{c}_{j}\hat{c}_{i}+\delta_{hb}\hat{c}_{c}\hat{c}_{a}\hat{c}_{j}\hat{c}_{i}\right)\\
 &   -\frac{L^{4}}{4}G_{ab,c}G_{ij,h}\left(\delta_{ic}\hat{c}_{b}\hat{c}_{a}\hat{c}_{h}\hat{c}_{j}+\delta_{jc}\hat{c}_{b}\hat{c}_{a}\hat{c}_{h}\hat{c}_{i}\right)\\
 &   +\frac{L^{4}}{4}G_{ab,c}G_{ij,h}\left(\delta_{hc}\hat{c}_{b}\hat{c}_{a}\hat{c}_{j}\hat{c}_{i}\right)\\
 &   -\frac{L^{4}}{4}G_{ab,c}G_{ij,h}\left(\hat{c}_{i}\hat{c}_{j}\hat{c}_{h}\hat{c}_{a}\hat{c}_{b}\hat{c}_{c}\right).
\end{align*}
Collecting the terms we have a new form of the squared tangent displacement,
\begin{align}
u_{\tau}^{2} &=   L^{2}G_{ab}G_{ij}\left(\delta_{ia}\hat{c}_{b}\hat{c}_{j}+\delta_{ja}\hat{c}_{b}\hat{c}_{i}+\delta_{ib}\hat{c}_{a}\hat{c}_{j}+\delta_{jb}\hat{c}_{a}\hat{c}_{i}-4\hat{c}_{i}\hat{c}_{j}\hat{c}_{a}\hat{c}_{b}\right)\label{eq: squared tangent displacement final form}\\
 &   +\frac{L^{3}}{2}G_{ab}G_{ij,h}\left(2\delta_{ia}\hat{c}_{b}\hat{c}_{h}\hat{c}_{j}+2\delta_{ja}\hat{c}_{b}\hat{c}_{h}\hat{c}_{i}-2\delta_{ha}\hat{c}_{b}\hat{c}_{j}\hat{c}_{i}+2\delta_{ib}\hat{c}_{a}\hat{c}_{h}\hat{c}_{j}\right)\nonumber \\
 &   +\frac{L^{3}}{2}G_{ab}G_{ij,h}\left(2\delta_{jb}\hat{c}_{a}\hat{c}_{h}\hat{c}_{i}-2\delta_{hb}\hat{c}_{a}\hat{c}_{j}\hat{c}_{i}-4\hat{c}_{i}\hat{c}_{j}\hat{c}_{a}\hat{c}_{b}\hat{c}_{c}\right)\nonumber \\
 &   +\frac{L^{4}}{4}G_{ab,c}G_{ij,h}\left(\delta_{ia}\hat{c}_{c}\hat{c}_{b}\hat{c}_{h}\hat{c}_{j}+\delta_{ja}\hat{c}_{c}\hat{c}_{b}\hat{c}_{h}\hat{c}_{i}-\delta_{ha}\hat{c}_{c}\hat{c}_{b}\hat{c}_{j}\hat{c}_{i}\right)\nonumber \\
 &   +\frac{L^{4}}{4}G_{ab,c}G_{ij,h}\left(\delta_{ib}\hat{c}_{c}\hat{c}_{a}\hat{c}_{h}\hat{c}_{j}+\delta_{jb}\hat{c}_{c}\hat{c}_{a}\hat{c}_{h}\hat{c}_{i}-\delta_{hb}\hat{c}_{c}\hat{c}_{a}\hat{c}_{j}\hat{c}_{i}\right)\nonumber \\
 &   -\frac{L^{4}}{4}G_{ab,c}G_{ij,h}\left(\delta_{ic}\hat{c}_{b}\hat{c}_{a}\hat{c}_{h}\hat{c}_{j}+\delta_{jc}\hat{c}_{b}\hat{c}_{a}\hat{c}_{h}\hat{c}_{i}-\delta_{hc}\hat{c}_{b}\hat{c}_{a}\hat{c}_{j}\hat{c}_{i}\right)\nonumber \\
 &   -\frac{L^{4}}{4}G_{ab,c}G_{ij,h}\left(\hat{c}_{i}\hat{c}_{j}\hat{c}_{h}\hat{c}_{a}\hat{c}_{b}\hat{c}_{c}\right).\nonumber 
\end{align}

\section{The density of the strain energy}

The density of the (internal) strain energy is, according to (15)
of \cite{barchiesietlZAMM2021}, the integral of the energy of each
orientation $\hat{c}$ over the unit circle $S^{\mathrm{1}}$ in the
2D case or over the unit sphere $S^{2}$ in the 3D case,
\begin{equation}
U=\int_{\mathcal{S^{\mathrm{1,2}}}}\left(\frac{1}{2}k_{\eta}u_{\eta}^{2}+\frac{1}{2}k_{\tau}u_{\tau}^{2}\right)dS,\label{eq:density elastic energy}
\end{equation}
where $k_{\eta}$ is the normal stiffness relative to the orientation
$\hat{c}$, $k_{\tau}$ is the tangent stiffness relative to the orientation
$\hat{c}$, and $dS$ is the angular element over $S^{\mathrm{1}}$
or $S^{2}$. Insertion of (\ref{eq:squared noormal displacement})
and (\ref{eq: squared tangent displacement final form}) into (\ref{eq:density elastic energy})
yields
\begin{align*}
U=\int_{\mathcal{S^{\mathrm{1,2}}}}\{ &  \frac{1}{2}k_{\eta}L^{2}G_{ij}G_{ab}\hat{c}_{i}\hat{c}_{j}\hat{c}_{a}\hat{c}_{b}\\
 &   +\frac{1}{2}k_{\eta}\frac{L^{3}}{2}G_{ij}G_{ab,c}\hat{c}_{i}\hat{c}_{j}\hat{c}_{a}\hat{c}_{b}\hat{c}_{c}\\
 &   +\frac{1}{2}k_{\eta}\frac{L^{4}}{16}G_{ij,h}G_{ab,c}\hat{c}_{i}\hat{c}_{j}\hat{c}_{h}\hat{c}_{a}\hat{c}_{b}\hat{c}_{c}\\
 &   \frac{1}{2}k_{\tau}L^{2}G_{ab}G_{ij}\left(\delta_{ia}\hat{c}_{b}\hat{c}_{j}+\delta_{ja}\hat{c}_{b}\hat{c}_{i}+\delta_{ib}\hat{c}_{a}\hat{c}_{j}+\delta_{jb}\hat{c}_{a}\hat{c}_{i}-4\hat{c}_{i}\hat{c}_{j}\hat{c}_{a}\hat{c}_{b}\right)\\
 &   +\frac{1}{2}k_{\tau}\frac{L^{3}}{2}G_{ab}G_{ij,h}\left(2\delta_{ia}\hat{c}_{b}\hat{c}_{h}\hat{c}_{j}+2\delta_{ja}\hat{c}_{b}\hat{c}_{h}\hat{c}_{i}-2\delta_{ha}\hat{c}_{b}\hat{c}_{j}\hat{c}_{i}+2\delta_{ib}\hat{c}_{a}\hat{c}_{h}\hat{c}_{j}\right)\\
 &   +\frac{1}{2}k_{\tau}\frac{L^{3}}{2}G_{ab}G_{ij,h}\left(2\delta_{jb}\hat{c}_{a}\hat{c}_{h}\hat{c}_{i}-2\delta_{hb}\hat{c}_{a}\hat{c}_{j}\hat{c}_{i}-4\hat{c}_{i}\hat{c}_{j}\hat{c}_{a}\hat{c}_{b}\hat{c}_{c}\right)\\
 &   +\frac{1}{2}k_{\tau}\frac{L^{4}}{4}G_{ab,c}G_{ij,h}\left(\delta_{ia}\hat{c}_{c}\hat{c}_{b}\hat{c}_{h}\hat{c}_{j}+\delta_{ja}\hat{c}_{c}\hat{c}_{b}\hat{c}_{h}\hat{c}_{i}-\delta_{ha}\hat{c}_{c}\hat{c}_{b}\hat{c}_{j}\hat{c}_{i}\right)\\
 &   +\frac{1}{2}k_{\tau}\frac{L^{4}}{4}G_{ab,c}G_{ij,h}\left(\delta_{ib}\hat{c}_{c}\hat{c}_{a}\hat{c}_{h}\hat{c}_{j}+\delta_{jb}\hat{c}_{c}\hat{c}_{a}\hat{c}_{h}\hat{c}_{i}-\delta_{hb}\hat{c}_{c}\hat{c}_{a}\hat{c}_{j}\hat{c}_{i}\right)\\
 &   -\frac{1}{2}k_{\tau}\frac{L^{4}}{4}G_{ab,c}G_{ij,h}\left(\delta_{ic}\hat{c}_{b}\hat{c}_{a}\hat{c}_{h}\hat{c}_{j}+\delta_{jc}\hat{c}_{b}\hat{c}_{a}\hat{c}_{h}\hat{c}_{i}-\delta_{hc}\hat{c}_{b}\hat{c}_{a}\hat{c}_{j}\hat{c}_{i}\right)\\
 &   -\frac{1}{2}k_{\tau}\frac{L^{4}}{4}G_{ab,c}G_{ij,h}\left(\hat{c}_{i}\hat{c}_{j}\hat{c}_{h}\hat{c}_{a}\hat{c}_{b}\hat{c}_{c}\right)\}dS.
\end{align*}
Collecting the terms we have an important form of the density of the
strain energy,
\begin{align*}
U=\frac{1}{2}G_{ab}G_{ij}L^{2}\int_{\mathcal{S^{\mathrm{1,2}}}}\{ &   \left(k_{\eta}-4k_{\tau}\right)\hat{c}_{i}\hat{c}_{j}\hat{c}_{a}\hat{c}_{b}\\
 &   k_{\tau}\left(\delta_{ia}\hat{c}_{b}\hat{c}_{j}+\delta_{ja}\hat{c}_{b}\hat{c}_{i}+\delta_{ib}\hat{c}_{a}\hat{c}_{j}+\delta_{jb}\hat{c}_{a}\hat{c}_{i}\right)\}dS\\
+G_{ab}G_{ij,h}\frac{L^{3}}{4}\int_{\mathcal{S^{\mathrm{1,2}}}}\{ &   \left(k_{\eta}-4k_{\tau}\right)\hat{c}_{a}\hat{c}_{b}\hat{c}_{i}\hat{c}_{j}\hat{c}_{h}\\
 &   +2k_{\tau}\left(\delta_{ia}\hat{c}_{b}\hat{c}_{h}\hat{c}_{j}+\delta_{ja}\hat{c}_{b}\hat{c}_{h}\hat{c}_{i}-\delta_{ha}\hat{c}_{b}\hat{c}_{j}\hat{c}_{i}\right)\\
 &   +2k_{\tau}\left(\delta_{ib}\hat{c}_{a}\hat{c}_{h}\hat{c}_{j}+\delta_{jb}\hat{c}_{a}\hat{c}_{h}\hat{c}_{i}-\delta_{hb}\hat{c}_{a}\hat{c}_{j}\hat{c}_{i}\right)\\
 &   \}dS\\
+\frac{1}{2}G_{ij,h}G_{ab,c}\frac{L^{4}}{16}\int_{\mathcal{S^{\mathrm{1,2}}}}\{   & \left(k_{\eta}-4k_{\tau}\right)\hat{c}_{i}\hat{c}_{j}\hat{c}_{h}\hat{c}_{a}\hat{c}_{b}\hat{c}_{c}\\
   & +4k_{\tau}\left(\delta_{ia}\hat{c}_{c}\hat{c}_{b}\hat{c}_{h}\hat{c}_{j}+\delta_{ja}\hat{c}_{c}\hat{c}_{b}\hat{c}_{h}\hat{c}_{i}-\delta_{ha}\hat{c}_{c}\hat{c}_{b}\hat{c}_{j}\hat{c}_{i}\right)\\
   & +4k_{\tau}\left(\delta_{ib}\hat{c}_{c}\hat{c}_{a}\hat{c}_{h}\hat{c}_{j}+\delta_{jb}\hat{c}_{c}\hat{c}_{a}\hat{c}_{h}\hat{c}_{i}-\delta_{hb}\hat{c}_{c}\hat{c}_{a}\hat{c}_{j}\hat{c}_{i}\right)\\
   & -4k_{\tau}\left(\delta_{ic}\hat{c}_{b}\hat{c}_{a}\hat{c}_{h}\hat{c}_{j}+\delta_{jc}\hat{c}_{b}\hat{c}_{a}\hat{c}_{h}\hat{c}_{i}-\delta_{hc}\hat{c}_{b}\hat{c}_{a}\hat{c}_{j}\hat{c}_{i}\right)\}dS,
\end{align*}
where we have taken into account the fact that the strain and strain
tensor components do not depend on the orientation $\hat{c}$ and
go out of the integral over the unit circle $S^{\mathrm{1}}$ in the
2D case or over the unit sphere $S^{2}$ in the 3D case. If we write
the previous expression in a standard form with the use of the stiffness
tensors $\mathbb{C}$, $\mathbb{M}$ and $\mathbb{D}$, 
\[
U=\frac{1}{2}\mathbb{C}_{abij}G_{ij}G_{ab}+\mathbb{M}_{abijh}G_{ab}G_{ij,h}+\frac{1}{2}\mathbb{D}_{abcijh}G_{ij,h}G_{ab,c}
\]
then we have the following identifications of the three stiffness
tensors,
\begin{align}
\mathbb{C}_{abij}=L^{2}\int_{\mathcal{S^{\mathrm{1,2}}}}\{ &  \left(k_{\eta}-4k_{\tau}\right)\hat{c}_{i}\hat{c}_{j}\hat{c}_{a}\hat{c}_{b}\label{eq:stiffness identification C}\\
 & k_{\tau}\left(\delta_{ia}\hat{c}_{b}\hat{c}_{j}+\delta_{ja}\hat{c}_{b}\hat{c}_{i}+\delta_{ib}\hat{c}_{a}\hat{c}_{j}+\delta_{jb}\hat{c}_{a}\hat{c}_{i}\right)\}dS,\nonumber 
\end{align}
\begin{align}
\mathbb{M}_{abijh}=\frac{L^{3}}{4}\int_{\mathcal{S^{\mathrm{1,2}}}}\{ &  \left(k_{\eta}-4k_{\tau}\right)\hat{c}_{a}\hat{c}_{b}\hat{c}_{i}\hat{c}_{j}\hat{c}_{h}\label{eq:eq:stiffness identification M}\\
 &   +2k_{\tau}\left(\delta_{ia}\hat{c}_{b}\hat{c}_{h}\hat{c}_{j}+\delta_{ja}\hat{c}_{b}\hat{c}_{h}\hat{c}_{i}-\delta_{ha}\hat{c}_{b}\hat{c}_{j}\hat{c}_{i}\right)\nonumber \\
 &   +2k_{\tau}\left(\delta_{ib}\hat{c}_{a}\hat{c}_{h}\hat{c}_{j}+\delta_{jb}\hat{c}_{a}\hat{c}_{h}\hat{c}_{i}-\delta_{hb}\hat{c}_{a}\hat{c}_{j}\hat{c}_{i}\right)\}dS,\nonumber 
\end{align}
\begin{align}
\mathbb{D}_{abcijh}=\frac{L^{4}}{16}\int_{\mathcal{S^{\mathrm{1,2}}}} &   \left(k_{\eta}-4k_{\tau}\right)\hat{c}_{i}\hat{c}_{j}\hat{c}_{h}\hat{c}_{a}\hat{c}_{b}\hat{c}_{c}\label{eq:eq:stiffness identification D}\\
 &   +4k_{\tau}\left(\delta_{ia}\hat{c}_{c}\hat{c}_{b}\hat{c}_{h}\hat{c}_{j}+\delta_{ja}\hat{c}_{c}\hat{c}_{b}\hat{c}_{h}\hat{c}_{i}-\delta_{ha}\hat{c}_{c}\hat{c}_{b}\hat{c}_{j}\hat{c}_{i}\right)\nonumber \\
 &   +4k_{\tau}\left(\delta_{ib}\hat{c}_{c}\hat{c}_{a}\hat{c}_{h}\hat{c}_{j}+\delta_{jb}\hat{c}_{c}\hat{c}_{a}\hat{c}_{h}\hat{c}_{i}-\delta_{hb}\hat{c}_{c}\hat{c}_{a}\hat{c}_{j}\hat{c}_{i}\right)\nonumber \\
 &   -4k_{\tau}\left(\delta_{ic}\hat{c}_{b}\hat{c}_{a}\hat{c}_{h}\hat{c}_{j}+\delta_{jc}\hat{c}_{b}\hat{c}_{a}\hat{c}_{h}\hat{c}_{i}-\delta_{hc}\hat{c}_{b}\hat{c}_{a}\hat{c}_{j}\hat{c}_{i}\right)\}dS.\nonumber 
\end{align}
A more synthetic way to propose such an identification is the following
\begin{equation}
\mathbb{C}_{abij}=L^{2}\int_{\mathcal{S^{\mathrm{1,2}}}}\left[\left(k_{\eta}-4k_{\tau}\right)\hat{c}_{i}\hat{c}_{j}\hat{c}_{a}\hat{c}_{b}+4k_{\tau}\delta_{(i(a}\hat{c}_{b)}\hat{c}_{j)}\right]dS,\label{eq:stiffness identification C shynthetic}
\end{equation}
\begin{equation}
\mathbb{M}_{abijh}=\frac{L^{3}}{4}\int_{\mathcal{S^{\mathrm{1,2}}}}\left[\left(k_{\eta}-4k_{\tau}\right)\hat{c}_{a}\hat{c}_{b}\hat{c}_{i}\hat{c}_{j}\hat{c}_{h}+4k_{\tau}\left(2\delta_{(i(a}\hat{c}_{b)}\hat{c}_{h}\hat{c}_{j)}-\delta_{h(a}\hat{c}_{b)}\hat{c}_{j}\hat{c}_{i}\right)\right]dS,\label{eq:stiffness identification M shynthetic}
\end{equation}
\begin{eqnarray}
 &  & \mathbb{D}_{abcijh}=\frac{L^{4}}{16}\int_{\mathcal{S^{\mathrm{1,2}}}}\{\left(k_{\eta}-4k_{\tau}\right)\hat{c}_{i}\hat{c}_{j}\hat{c}_{h}\hat{c}_{a}\hat{c}_{b}\hat{c}_{c}\label{eq:stiffness identification D shynthetic}\\
 &  & +4k_{\tau}\left(4\delta_{(i(a}\hat{c}_{c}\hat{c}_{b)}\hat{c}_{h}\hat{c}_{j)}-\delta_{h(a}\hat{c}_{c}\hat{c}_{b)}\hat{c}_{j}\hat{c}_{i}-2\delta_{(ic}\hat{c}_{b}\hat{c}_{a}\hat{c}_{h}\hat{c}_{j)}+\delta_{hc}\hat{c}_{b}\hat{c}_{a}\hat{c}_{j}\hat{c}_{i}\right)\}dS,\nonumber 
\end{eqnarray}
where the parentheses on the index means the standard single symmetrization
rule
\[
A_{\left(iklj\right)}=\frac{1}{2}\left(A_{iklj}+A_{jkli}\right),
\]
and nested parentheses are symmetrized first,
\[
A_{(a(bc)d)}=\frac{1}{2}\left(A_{(abcd)}+A_{(acbd)}\right)=\frac{1}{4}\left(A_{abcd}+A_{acbd}+A_{dbca}+A_{dcba}\right)
\]
It is worth to note that the identification (\ref{eq:stiffness identification C})
is the same as equation (18) of \cite{barchiesietlZAMM2021}. However, the identifications
(\ref{eq:eq:stiffness identification M}-\ref{eq:eq:stiffness identification D})
change with respect to equation (19) of \cite{barchiesietlZAMM2021}. This change is the
aim of this paper.

\section{Identification of the 2D isotropic case}

In the general anisotropic case, normal $k_{\eta}$ and tangent $k_{\tau}$
stiffnesses are functions of the orientation $\hat{c}$. However,
in the isotropic case we consider no functional dependence. In the
2D case, where the integration of (\ref{eq:stiffness identification C}-\ref{eq:eq:stiffness identification M}-\ref{eq:eq:stiffness identification D})
or (\ref{eq:stiffness identification C shynthetic}-\ref{eq:stiffness identification M shynthetic}-\ref{eq:stiffness identification D shynthetic})
is done over the unit circle $S^{\mathrm{1}}$, the equation that
follows equation (36) of \cite{barchiesietlZAMM2021} is,
\begin{equation}
k_{\eta}=\frac{\bar{k}_{\eta}}{2\pi},\qquad k_{\tau}=\frac{\bar{k}_{\tau}}{2\pi},\label{eq:iso2D}
\end{equation}
where $\bar{k}_{\eta}$ and $\bar{k}_{\tau}$ are the integrated stiffnesses
over the set of possible orientations. An analytical calculation of
the insertion of (\ref{eq:iso2D}) into (\ref{eq:stiffness identification C shynthetic})
yields the following identification of the non null components of
the standard stiffness tensor. They are divided into 3 groups. 

\begin{eqnarray*}
 &  & \mathbb{C}_{1111}=\mathbb{C}_{2222}=\frac{1}{8}L^{2}\left(3\bar{k}_{\eta}+4\bar{k}_{\tau}\right),\\
 &  & \mathbb{C}_{1122}=\mathbb{C}_{2211}=\frac{1}{8}L^{2}\left(\bar{k}_{\eta}-4\bar{k}_{\tau}\right),\\
 &  & \mathbb{C}_{1212}=\mathbb{C}_{1221}=\mathbb{C}_{2112}=\mathbb{C}_{2121}=\frac{1}{8}L^{2}\left(\bar{k}_{\eta}+4\bar{k}_{\tau}\right).
\end{eqnarray*}
The rest of the components are null. The identification of the Lam\'e
parameters are
\[
\lambda=\mathbb{C}_{1122}=\frac{1}{8}L^{2}\left(\bar{k}_{\eta}-4\bar{k}_{\tau}\right),\;\mu=\mathbb{C}_{1212}=\frac{L^{2}}{8}\left(\bar{k}_{\eta}+4\bar{k}_{\tau}\right)=\frac{1}{2}\left(\mathbb{C}_{1111}-\mathbb{C}_{1122}\right).
\]
The Young's modulus $Y_{2D}$ and Poisson's ratio $\nu_{2D}$ are
given from the expressions (A.12) of \cite{barchiesietlZAMM2021},
\begin{equation}
Y_{2D}=4\mu\frac{\lambda+\mu}{\lambda+2\mu}=\frac{L^{2}\bar{k}_{\eta}\left(\bar{k}_{\eta}+4\bar{k}_{\tau}\right)}{3\bar{k}_{\eta}+4\bar{k}_{\tau}},\qquad\nu_{2D}=\frac{\lambda}{\lambda+2\mu}=\frac{\bar{k}_{\eta}-4\bar{k}_{\tau}}{3\bar{k}_{\eta}+4\bar{k}_{\tau}}.\label{eq:Y2dnu2d}
\end{equation}
 The inversion of (\ref{eq:Y2dnu2d}) gives the normal and tangent
integrated stiffnesses in terms of both the Young's modulus $Y_{2D}$
and the Poisson's ratio $\nu_{2D}$,
\begin{equation}
\bar{k}_{\eta}=-\frac{2Y_{2D}}{L^{2}\left(\nu_{2D}-1\right)},\qquad\bar{k}_{\tau}=\frac{\left(3\nu_{2D}-1\right)Y_{2D}}{2L^{2}\left(\nu_{2D}^{2}-1\right)}.\label{eq:ketaktau2Dinterms of youngnu}
\end{equation}

An analytical calculation of the insertion of (\ref{eq:iso2D}) into
(\ref{eq:eq:stiffness identification D}) yields the following identification
of the non-null components of the strain gradient stiffness tensor.
They are divided into 6 groups. 

The first,
\[
d_{1}=\mathbb{D}_{111111}=\mathbb{D}_{222222},
\]
that, in terms of integrated stiffnesses $\bar{k}_{\eta}$ and $\bar{k}_{\tau}$
or in terms of the Young's modulus $Y_{2D}$ and Poisson's ratio $\nu_{2D}$,
is
\[
d_{1}=\frac{L^{4}}{256}\left(5\bar{k}_{\eta}+4\bar{k}_{\tau}\right)=\frac{L^{2}\left(3+\nu_{2D}\right)Y_{2D}}{64\left(1-\nu_{2D}^{2}\right)}.
\]

The second,
\begin{eqnarray*}
 &  & d_{2}=\mathbb{D}_{111122}=\mathbb{D}_{111212}=\mathbb{D}_{121222}=\mathbb{D}_{122111},\\
 &  & d_{2}=\mathbb{D}_{211222}=\mathbb{D}_{212111}=\mathbb{D}_{222121}=\mathbb{D}_{222211},
\end{eqnarray*}
that, in terms of integrated stiffnesses $\bar{k}_{\eta}$ and $\bar{k}_{\tau}$
or in terms of the Young's modulus $Y_{2D}$ and Poisson's ratio $\nu_{2D}$,
is
\[
d_{2}=\frac{L^{4}}{256}\left(\bar{k}_{\eta}+4\bar{k}_{\tau}\right)=\frac{L^{2}Y_{2D}}{64\left(\nu_{2D}+1\right)}.
\]

The third,
\[
d_{3}=\mathbb{D}_{111221}=\mathbb{D}_{112222}=\mathbb{D}_{221111}=\mathbb{D}_{222112},
\]
that ,in terms of integrated stiffnesses $\bar{k}_{\eta}$ and $\bar{k}_{\tau}$
or in terms of the Young's modulus $Y_{2D}$ and Poisson's ratio $\nu_{2D}$,
is
\[
d_{3}=\frac{L^{4}}{256}\left(\bar{k}_{\eta}-12\bar{k}_{\tau}\right)=\frac{L^{2}\left(1-5\nu_{2D}\right)Y_{2D}}{64\left(\nu_{2D}^{2}-1\right)}.
\]

The fourth,
\begin{eqnarray*}
 &  & d_{4}=\mathbb{D}_{112112}=\mathbb{D}_{221221},
\end{eqnarray*}
that, in terms of integrated stiffnesses $\bar{k}_{\eta}$ and $\bar{k}_{\tau}$
or in terms of the Young's modulus $Y_{2D}$ and Poisson's ratio $\nu_{2D}$,
is
\[
d_{4}=\frac{L^{4}}{256}\left(\bar{k}_{\eta}+52\bar{k}_{\tau}\right)=\frac{L^{2}\left(19\nu_{2D}-7\right)Y_{2D}}{64\left(\nu_{2D}^{2}-1\right)}.
\]

The fifth,
\begin{eqnarray*}
 &  & d_{5}=\mathbb{D}_{112121}=\mathbb{D}_{112211}=\mathbb{D}_{121112}=\mathbb{D}_{122221},\\
 &  & d_{5}=\mathbb{D}_{211112}=\mathbb{D}_{212221}=\mathbb{D}_{221122}=\mathbb{D}_{221212},
\end{eqnarray*}
that, in terms of integrated stiffness $\bar{k}_{\eta}$ and $\bar{k}_{\tau}$
or in terms of the Young's modulus $Y_{2D}$ and Poisson's ratio $\nu_{2D}$,
is
\[
d_{5}=\frac{L^{4}}{256}\left(\bar{k}_{\eta}-28\bar{k}_{\tau}\right)=\frac{L^{2}\left(3-11\nu_{2D}\right)Y_{2D}}{64\left(\nu_{2D}^{2}-1\right)}.
\]

The sixth,
\begin{eqnarray*}
 &  & d_{6}=\mathbb{D}_{121121}=\mathbb{D}_{121211}=\mathbb{D}_{122122}=\mathbb{D}_{122212},\\
 &  & d_{6}=\mathbb{D}_{211121}=\mathbb{D}_{211211}=\mathbb{D}_{212122}=\mathbb{D}_{212212},
\end{eqnarray*}
that, in terms of integrated stiffness $\bar{k}_{\eta}$ and $\bar{k}_{\tau}$
or in terms of the Young's modulus $Y_{2D}$ and Poisson's ratio $\nu_{2D}$,
is
\[
d_{6}=\frac{L^{4}}{256}\left(\bar{k}_{\eta}+20\bar{k}_{\tau}\right)=\frac{L^{2}\left(7\nu_{2D}-3\right)Y_{2D}}{64\left(\nu_{2D}^{2}-1\right)}.
\]

The rest of the components of the stiffness tensor are null.

\section{Identification of the 3D isotropic case}

In the general anisotropic case, normal $k_{\eta}$ and tangent $k_{\tau}$
stiffnesses are functions of the orientation $\hat{c}$. However,
in the isotropic case we consider no such dependence. In the 3D case,
where the integration of (\ref{eq:stiffness identification C}-\ref{eq:eq:stiffness identification M}-\ref{eq:eq:stiffness identification D})
or (\ref{eq:stiffness identification C shynthetic}-\ref{eq:stiffness identification M shynthetic}-\ref{eq:stiffness identification D shynthetic})
is done over the unit sphere $S^{\mathrm{2}}$, the expression of
the equation that follows equation (56) of \cite{barchiesietlZAMM2021}
is,

\begin{equation}
k_{\eta}=\frac{\bar{k}_{\eta}}{4\pi},\qquad k_{\tau}=\frac{\bar{k}_{\tau}}{4\pi},\label{eq:iso3D}
\end{equation}
where $\bar{k}_{\eta}$ and $\bar{k}_{\tau}$ are the are integrated
stiffnesses over the set of possible orientations. An analytical calculation
of the insertion of (\ref{eq:iso3D}) into (\ref{eq:stiffness identification C shynthetic})
yields the following identification of the non-null components of
the standard stiffness tensor. They are divided into 3 groups. 
\[
\mathbb{C}_{1111}=\mathbb{C}_{2222}=\mathbb{C}_{3333}=\frac{L^{2}}{15}\left(3\bar{k}_{\eta}+8\bar{k}_{\tau}\right),
\]
\[
\mathbb{C}_{1122}=\mathbb{C}_{1133}=\mathbb{C}_{2211}=\mathbb{C}_{2233}=\mathbb{C}_{3311}=\mathbb{C}_{3322}=\frac{L^{2}}{15}\left(\bar{k}_{\eta}-4\bar{k}_{\tau}\right),
\]
\begin{eqnarray*}
 &  & \mathbb{C}_{1212}=\mathbb{C}_{1221}=\mathbb{C}_{1313}=\mathbb{C}_{1331}=\mathbb{C}_{2112}=\mathbb{C}_{2121}=\frac{L^{2}}{15}\left(\bar{k}_{\eta}+6\bar{k}_{\tau}\right),\\
 &  & \mathbb{C}_{2323}=\mathbb{C}_{2332}=\mathbb{C}_{3113}=\mathbb{C}_{3131}=\mathbb{C}_{3223}=\mathbb{C}_{3232}=\frac{L^{2}}{15}\left(\bar{k}_{\eta}+6\bar{k}_{\tau}\right).
\end{eqnarray*}
The rest of the components are null. Identification of the Lam\'e parameters
are
\[
\lambda=\mathbb{C}_{1122}=\frac{1}{15}L^{2}\left(\bar{k}_{\eta}-4\bar{k}_{\tau}\right),\;\mu=\mathbb{C}_{1212}=\frac{L^{2}}{8}\left(\bar{k}_{\eta}+6\bar{k}_{\tau}\right)=\frac{1}{2}\left(\mathbb{C}_{1111}-\mathbb{C}_{1122}\right).
\]
The Young's modulus $Y_{3D}$ and Poisson's ratio $\nu_{3D}$ are
given from the expressions (A.15) of \cite{barchiesietlZAMM2021},
\begin{equation}
Y_{3D}=\mu\frac{3\lambda+\mu}{\lambda+2\mu}=\frac{L^{2}\bar{k}_{\eta}\left(\bar{k}_{\eta}+6\bar{k}_{\tau}\right)}{6\left(\bar{k}_{\eta}+\bar{k}_{\tau}\right)},\qquad\nu_{3D}=\frac{\lambda}{2\left(\lambda+\mu\right)}=\frac{\bar{k}_{\eta}-4\bar{k}_{\tau}}{4\left(\bar{k}_{\eta}+\bar{k}_{\tau}\right)}.\label{eq:Y3dnu3d}
\end{equation}
The inversion of (\ref{eq:Y3dnu3d}) gives the normal and tangent
integrated stiffnesses in terms of the Young's modulus $Y_{3D}$ and
the Poisson's ratio $\nu_{3D}$,
\begin{equation}
\bar{k}_{\eta}=\frac{3Y_{3D}}{L^{2}\left(1-2\nu_{3D}\right)},\qquad\bar{k}_{\tau}=\frac{\left(4\nu_{3D}-1\right)3Y_{3D}}{4L^{2}\left(2\nu_{3D}^{2}+\nu_{3D}-1\right)}.\label{eq:ketaktau3Dinterms of youngnu}
\end{equation}
If we represent the strain gradient stiffness matrix with the following
isotropic representation
\begin{eqnarray}
 &  & \mathbb{D}_{ijklmn}=c_{3}\left(\delta_{ij}\delta_{kl}\delta_{mn}+\delta_{in}\delta_{jk}\delta_{lm}+\delta_{ij}\delta_{km}\delta_{ln}+\delta_{ik}\delta_{jn}\delta_{lm}\right)\label{eq: Iso Dijklmn}\\
 &  & +c_{4}\delta_{ij}\delta_{kn}\delta_{ml}\nonumber \\
 &  & +c_{5}\left(\delta_{ik}\delta_{jl}\delta_{mn}+\delta_{im}\delta_{jk}\delta_{ln}+\delta_{ik}\delta_{jm}\delta_{ln}+\delta_{il}\delta_{jk}\delta_{mn}\right)\nonumber \\
 &  & +c_{6}\left(\delta_{il}\delta_{jm}\delta_{kn}+\delta_{im}\delta_{jl}\delta_{kn}\right)\nonumber \\
 &  & +c_{7}\left(\delta_{il}\delta_{jn}\delta_{mk}+\delta_{im}\delta_{jn}\delta_{lk}+\delta_{in}\delta_{jl}\delta_{km}+\delta_{in}\delta_{jm}\delta_{kl}\right)\nonumber 
\end{eqnarray}
then, it is sufficient to know the following five components of $\mathbb{D}$,
\begin{eqnarray}
 &  & \mathbb{D}_{111111}=4c_{3}+c_{4}+4c_{5}+2c_{6}+4c_{7},\quad\mathbb{D}_{221221}=c_{4}+2c_{6},\label{eq: 5indcomp1}\\
 &  & \mathbb{D}_{111221}=2c_{3}+c_{4},\quad\mathbb{D}_{221122}=c_{3}+2c_{7},\quad\mathbb{D}_{112233}=c_{3},\label{eq: 5indcomp2}
\end{eqnarray}
to identify the five independent isotropic strain gradient coefficients
as follows,
\begin{eqnarray}
 &  & c_{3}=\mathbb{D}_{112233},\quad c_{4}=\mathbb{D}_{111221}-2\mathbb{D}_{112233},\label{eq:cD1}\\
 &  & c_{5}=\frac{1}{4}\left(\mathbb{D}_{111111}-2\mathbb{D}_{112233}-2\mathbb{D}_{221122}-\mathbb{D}_{221221}\right),\label{eq:cD2}\\
 &  & c_{6}=\frac{1}{2}\left(-\mathbb{D}_{111221}+2\mathbb{D}_{112233}+\mathbb{D}_{221221}\right),\quad c_{7}=\frac{1}{2}\left(-\mathbb{D}_{112233}+\mathbb{D}_{221122}\right).\label{eq:cD3}
\end{eqnarray}
It is worth to note that the expressions (\ref{eq: 5indcomp1}-\ref{eq: 5indcomp2})
have been calculated from (\ref{eq: Iso Dijklmn}) and the identifications
(\ref{eq:cD1}-\ref{eq:cD2}) is given by the inversion of (\ref{eq: 5indcomp1}-\ref{eq: 5indcomp2}).
An analytical calculation of the insertion of (\ref{eq:iso3D}) into
(\ref{eq:eq:stiffness identification D}) yields the identification
of these 5 components of the strain gradient stiffness tensor,

\begin{eqnarray}
 &  & \mathbb{D}_{111111}=\frac{L^{4}}{560}\left(5\bar{k}_{\eta}+8\bar{k}_{\tau}\right)=\frac{3L^{2}\left(3\nu_{3D}-7\right)Y_{3D}}{560\left(2\nu_{3D}^{2}+\nu_{3D}-1\right)},\label{eq:D111111}\\
 &  & \mathbb{D}_{221221}=\frac{L^{4}\left(3\bar{k}_{\eta}+184\bar{k}_{\tau}\right)}{1680}=\frac{L^{2}\left(181\nu_{3D}-49\right)Y_{3D}}{560\left(2\nu_{3D}^{2}+\nu_{3D}-1\right)},\label{eq:D221221}\\
 &  & \mathbb{D}_{111221}=\frac{L^{4}\left(3\bar{k}_{\eta}-40\bar{k}_{\tau}\right)}{1680}=\frac{L^{2}\left(7-43\nu_{3D}\right)Y_{3D}}{560\left(2\nu_{3D}^{2}+\nu_{3D}-1\right)},\label{eq:D111221}\\
 &  & \mathbb{D}_{221122}=\frac{L^{4}\left(\bar{k}_{\eta}-32\bar{k}_{\tau}\right)}{560}=\frac{3L^{2}\left(7-33\nu_{3D}\right)Y_{3D}}{560\left(2\nu_{3D}^{2}+\nu_{3D}-1\right)},\label{eq:D221122}\\
 &  & \mathbb{D}_{112233}=\frac{L^{4}\left(\bar{k}_{\eta}-32\bar{k}_{\tau}\right)}{1680}=\frac{L^{2}\left(7-33\nu_{3D}\right)Y_{3D}}{560\left(2\nu_{3D}^{2}+\nu_{3D}-1\right)},\label{eq:D112233}
\end{eqnarray}
where the second equivalence is achieved because of (\ref{eq:ketaktau3Dinterms of youngnu}).
Insertion of (\ref{eq:D111111}-\ref{eq:D112233}) into (\ref{eq:cD1}-\ref{eq:cD2})
yields
\begin{eqnarray*}
 &  & c_{3}=\frac{L^{4}}{1680}\left(\bar{k}_{\eta}-32\bar{k}_{\tau}\right)=\frac{L^{2}\left(7-33\nu_{3D}\right)Y_{3D}}{560\left(2\nu_{3D}^{2}+\nu_{3D}-1\right)},\\
 &  & c_{4}=\frac{L^{4}}{1680}\left(\bar{k}_{\eta}+24\bar{k}_{\tau}\right)=\frac{L^{2}\left(23\nu_{3D}-7\right)Y_{3D}}{560\left(2\nu_{3D}^{2}+\nu_{3D}-1\right)},\\
 &  & c_{5}=\frac{L^{4}}{1680}\left(\bar{k}_{\eta}+24\bar{k}_{\tau}\right)=\frac{L^{2}\left(23\nu_{3D}-7\right)Y_{3D}}{560\left(2\nu_{3D}^{2}+\nu_{3D}-1\right)},\\
 &  & c_{6}=\frac{L^{4}}{1680}\left(\bar{k}_{\eta}+80\bar{k}_{\tau}\right)=\frac{L^{2}\left(79\nu_{3D}-21\right)Y_{3D}}{560\left(2\nu_{3D}^{2}+\nu_{3D}-1\right)},\\
 &  & c_{7}=\frac{L^{4}}{1680}\left(\bar{k}_{\eta}-32\bar{k}_{\tau}\right)=\frac{L^{2}\left(7-33\nu_{3D}\right)Y_{3D}}{560\left(2\nu_{3D}^{2}+\nu_{3D}-1\right)},
\end{eqnarray*}
where the second equivalence is achieved because of (\ref{eq:ketaktau3Dinterms of youngnu}). 

The Mindlin coefficients $a_{1}$, $a_{2}$, $a_{3}$, $a_{4}$ and
$a_{5}$ are then related to $c_{3}$, $c_{4}$, $c_{5}$, $c_{6}$
and $c_{7}$ via
\[
a_{1}=2c_{3},\quad a_{2}=2c_{4},\quad a_{3}=2c_{5},\quad a_{4}=c_{6},\quad a_{5}=2c_{7}.
\]

\section{Concluding Remarks}

The main contribution of this paper is a corrected expression for the grain-pair objective relative displacement within the granular micromechanics framework that incorporates second gradients of placement. This correction amends the formulation presented in \cite{barchiesietlZAMM2021} and leads to modified strain energy contributions from both the normal and tangential components of grain-pair interactions. As a result, the elastic energy density functional in the second gradient granular micromechanics approach is updated, and explicit stiffness tensors applicable to anisotropic second gradient elasticity are derived. We also revise the analytical expressions for the material parameters of 2D and 3D isotropic second gradient elasticity previously reported in \cite{barchiesietlZAMM2021}. The standard (first gradient) elastic parameters remain unchanged but are included here for completeness.

Finally, we note that second gradient elasticity is increasingly used to explain observed mechanical responses exhibiting intrinsic length scale effects. The parameters provided in this work offer a practical starting point for such applications, especially since the formulation reveals the emergence of multiple characteristic lengths under different deformation modes.

\section*{Appendix: Complete list of strain gradient stiffness components}

In this appendix, we list the complete list of the strain gradient
stiffness tensor components. They are divided into 7 groups.

The first group is 
\[
d_{1}=\mathbb{D}_{111111}=\mathbb{D}_{222222}=\mathbb{D}_{333333}=\frac{L^{4}}{560}\left(5\bar{k}_{\eta}+8\bar{k}_{\tau}\right).
\]

The second group is
\begin{eqnarray*}
d_{2}= &  & \mathbb{D}_{111122}=\mathbb{D}_{111133}=\mathbb{D}_{111212}=\mathbb{D}_{111313}=\mathbb{D}_{121222}=\mathbb{D}_{122111}=\mathbb{D}_{131333}=\mathbb{D}_{133111},\\
d_{2}= &  & \mathbb{D}_{211222}=\mathbb{D}_{212222}=\mathbb{D}_{222121}=\mathbb{D}_{222211}=\mathbb{D}_{222233}=\mathbb{D}_{222323}=\mathbb{D}_{232333}=\mathbb{D}_{233222},\\
d_{2}= &  & \mathbb{D}_{311333}=\mathbb{D}_{313111}=\mathbb{D}_{322333}=\mathbb{D}_{323222}=\mathbb{D}_{333131}=\mathbb{D}_{333232}=\mathbb{D}_{333311}=\mathbb{D}_{333322},
\end{eqnarray*}
where
\[
d_{2}=\frac{L^{4}}{1680}\left(3\bar{k}_{\eta}+16\bar{k}_{\tau}\right).
\]

The third group is
\begin{eqnarray*}
d_{3}= &  & \mathbb{D}_{111221}=\mathbb{D}_{111331}=\mathbb{D}_{112222}=\mathbb{D}_{113333},\\
d_{3}= &  & \mathbb{D}_{221111}=\mathbb{D}_{222112}=\mathbb{D}_{222332}=\mathbb{D}_{223333},\\
d_{3}= &  & \mathbb{D}_{331111}=\mathbb{D}_{332222}=\mathbb{D}_{333113}=\mathbb{D}_{333223},
\end{eqnarray*}
where
\[
d_{3}=\frac{L^{4}}{1680}\left(3\bar{k}_{\eta}-40\bar{k}_{\tau}\right).
\]

The fourth group is
\begin{eqnarray*}
d_{4}= &  & \mathbb{D}_{112112}=\mathbb{D}_{113113}=\mathbb{D}_{221221}=\mathbb{D}_{223223}=\mathbb{D}_{331331}=\mathbb{D}_{332332},
\end{eqnarray*}
where
\[
d_{4}=\frac{L^{4}}{1680}\left(3\bar{k}_{\eta}+184\bar{k}_{\tau}\right).
\]

The fifth group is
\begin{eqnarray*}
d_{5}= &  & \mathbb{D}_{112121}=\mathbb{D}_{112211}=\mathbb{D}_{113131}=\mathbb{D}_{113311}=\mathbb{D}_{121112}=\mathbb{D}_{122221}=\mathbb{D}_{131113}=\mathbb{D}_{133331},\\
d_{5}= &  & \mathbb{D}_{211112}=\mathbb{D}_{212221}=\mathbb{D}_{221122}=\mathbb{D}_{221212}=\mathbb{D}_{223232}=\mathbb{D}_{223322}=\mathbb{D}_{232223}=\mathbb{D}_{233332},\\
d_{5}= &  & \mathbb{D}_{311113}=\mathbb{D}_{313331}=\mathbb{D}_{322223}=\mathbb{D}_{323332}=\mathbb{D}_{331133}=\mathbb{D}_{331313}=\mathbb{D}_{332233}=\mathbb{D}_{332323},\\
3d_{5}= &  & \mathbb{D}_{112233}=\mathbb{D}_{112323}=\mathbb{D}_{113232}=\mathbb{D}_{113322}=\mathbb{D}_{121332}=\mathbb{D}_{122331}=\mathbb{D}_{123132}=\mathbb{D}_{123231},\\
3d_{5}= &  & \mathbb{D}_{123312}=\mathbb{D}_{123321}=\mathbb{D}_{131223}=\mathbb{D}_{132123}=\mathbb{D}_{132213}=\mathbb{D}_{132231}=\mathbb{D}_{132321}=\mathbb{D}_{133221},\\
3d_{5}= &  & \mathbb{D}_{211332}=\mathbb{D}_{212331}=\mathbb{D}_{213132}=\mathbb{D}_{213231}=\mathbb{D}_{213312}=\mathbb{D}_{213321}=\mathbb{D}_{221133}=\mathbb{D}_{221313},\\
3d_{5}= &  & \mathbb{D}_{223131}=\mathbb{D}_{223311}=\mathbb{D}_{231123}=\mathbb{D}_{231132}=\mathbb{D}_{231213}=\mathbb{D}_{231312}=\mathbb{D}_{232113}=\mathbb{D}_{233112},\\
3d_{5}= &  & \mathbb{D}_{311223}=\mathbb{D}_{312123}=\mathbb{D}_{312213}=\mathbb{D}_{312231}=\mathbb{D}_{312321}=\mathbb{D}_{313221}=\mathbb{D}_{321123}=\mathbb{D}_{321132},\\
3d_{5}= &  & \mathbb{D}_{321213}=\mathbb{D}_{321312}=\mathbb{D}_{322113}=\mathbb{D}_{323112}=\mathbb{D}_{331122}=\mathbb{D}_{331212}=\mathbb{D}_{332121}=\mathbb{D}_{332211},
\end{eqnarray*}
where
\[
d_{5}=\frac{L^{4}}{560}\left(\bar{k}_{\eta}-32\bar{k}_{\tau}\right).
\]

The sixth group is
\begin{eqnarray*}
d_{6}= &  & \mathbb{D}_{112332}=\mathbb{D}_{113223}=\mathbb{D}_{121233}=\mathbb{D}_{121323}=\mathbb{D}_{122133},\\
d_{6}= &  & \mathbb{D}_{122313}=\mathbb{D}_{131232}=\mathbb{D}_{131322}=\mathbb{D}_{133122}=\mathbb{D}_{133212},\\
d_{6}= &  & \mathbb{D}_{211233}=\mathbb{D}_{211323}=\mathbb{D}_{212133}=\mathbb{D}_{212313}=\mathbb{D}_{221331},\\
d_{6}= &  & \mathbb{D}_{223113}=\mathbb{D}_{232131}=\mathbb{D}_{232311}=\mathbb{D}_{233121}=\mathbb{D}_{233211},\\
d_{6}= &  & \mathbb{D}_{311232}=\mathbb{D}_{311322}=\mathbb{D}_{313122}=\mathbb{D}_{313212}=\mathbb{D}_{322131},\\
d_{6}= &  & \mathbb{D}_{322311}=\mathbb{D}_{323121}=\mathbb{D}_{323211}=\mathbb{D}_{331221}=\mathbb{D}_{332112},\\
3d_{6}= &  & \mathbb{D}_{121121}=\mathbb{D}_{121211}=\mathbb{D}_{122122}=\mathbb{D}_{122212}=\mathbb{D}_{131131}=\mathbb{D}_{131311}=\mathbb{D}_{133133}=\mathbb{D}_{133313},\\
3d_{6}= &  & \mathbb{D}_{211121}=\mathbb{D}_{211211}=\mathbb{D}_{212122}=\mathbb{D}_{212212}=\mathbb{D}_{232232}=\mathbb{D}_{232322}=\mathbb{D}_{233233}=\mathbb{D}_{233323},\\
3d_{6}= &  & \mathbb{D}_{311131}=\mathbb{D}_{311311}=\mathbb{D}_{313133}=\mathbb{D}_{313313}=\mathbb{D}_{322232}=\mathbb{D}_{322322}=\mathbb{D}_{323233}=\mathbb{D}_{323323},
\end{eqnarray*}
where
\[
d_{6}=\frac{L^{4}}{1680}\left(\bar{k}_{\eta}+24\bar{k}_{\tau}\right).
\]

The seventh group is
\begin{eqnarray*}
d_{7}= &  & \mathbb{D}_{123123}=\mathbb{D}_{123213}=\mathbb{D}_{132132}=\mathbb{D}_{132312},\\
d_{7}= &  & \mathbb{D}_{213123}=\mathbb{D}_{213213}=\mathbb{D}_{231231}=\mathbb{D}_{231321},\\
d_{7}= &  & \mathbb{D}_{312132}=\mathbb{D}_{312312}=\mathbb{D}_{321231}=\mathbb{D}_{321321},
\end{eqnarray*}
where
\[
d_{7}=\frac{L^{4}}{1680}\left(\bar{k}_{\eta}+80\bar{k}_{\tau}\right).
\]

\end{document}